\documentclass
[preprint,pra,showpacs,showkeys,byrevtex,nobibnotes,nofootinbib,10pt,letterpaper,onecolumn]{revtex4}%
\usepackage{amsfonts}
\usepackage{amsmath}
\usepackage{amssymb}
\usepackage{graphicx}%
\usepackage{epsfig}
\providecommand{\U}[1]{\protect\rule{.1in}{.1in}}
\begin{document}
\title{Decoherence: a closed-system approach}
\author{Sebastian Fortin}
\affiliation{CONICET, IAFE (CONICET-UBA) and FCEN (UBA), Argentina.}
\author{Olimpia Lombardi}
\affiliation{CONICET and FCEN (UBA), Argentina.}
\author{Mario Castagnino}
\affiliation{CONICET, IAFE (CONICET-UBA), IFIR and FCEN (UBA), Argentina.}

\begin{abstract}
The aim of this paper is to review a new perspective about decoherence,
according to which formalisms originally devised to deal just with closed or
open systems can be subsumed under a closed-system approach that generalizes
the traditional account of the phenomenon. This new viewpoint dissolves
certain conceptual difficulties of the orthodox open-system approach but, at
the same time, shows that the openness of the quantum system is not the
essential ingredient for decoherence, as commonly claimed. Moreover, when the
behavior of a decoherent system is described from a closed-system perspective,
the account of decoherence turns out to be more general than that supplied by
the open-system approach, and the quantum-to-classical transition defines
unequivocally the realm of classicality by identifying the observables with
classical-like behavior.

\end{abstract}
\maketitle

\section{\textbf{Introduction}}

Since more than two decades ago, the environment-induced decoherence (EID)
approach is considered a \textquotedblleft new orthodoxy\textquotedblright\ in
the physicists community (\cite{Leggett}, \cite{Bub}). It has been fruitfully
applied in many areas of physics and supplies the basis of new technological
developments. In the foundations of physics community, EID has been viewed as
a relevant element for the interpretation of quantum mechanics (\cite{BH-1994}%
, \cite{BH-1996}) and for the explanation of the emergence of classicality
from the quantum world (\cite{Elby}, \cite{Healey}, \cite{Paz-Zurek-2002},
\cite{Zurek-2003}).

The great success of EID has given rise to the idea that decoherence
necessarily requires the interaction between an open quantum system and an
environment of many, potentially infinite, degrees of freedom. However, the
historical roots of the decoherence program can be found in certain attempts
to explain the emergence of classicality in closed systems. In turn, at
present other approaches have been proposed, and in several of them the
openness of the system is not an essential factor. These new approaches are
usually conceived as rival to EID, or even as dealing with different physical phenomena.

On the basis of our previous works on the subject, the aim of this paper is to
supply a comprehensive presentation of a new perspective, according to which
formalisms originally devised to deal just with closed or open systems can be
subsumed under a closed-system approach that generalizes the traditional
account of the phenomenon. With this purpose, in Section II we will briefly
review the historical development of the decoherence program, and in Section
III we will stress some conceptual difficulties of the EID program. In Section
IV we will recall the orthodox approach to decoherence, emphasizing its
open-system character. Section V will be devoted to introduce a closed-system
perspective, and to use it to reformulate the orthodox open-system EID
approach. On this basis, in Section VI we will formulate the closed-system
approach from a generic viewpoint: in this framework, the account of
decoherence turns out to be more general than that supplied by the orthodox
open-system approach. In order to illustrate this theoretical claim, in
Section VII we will consider the description of the well-known spin-bath model
in the context of the closed-system framework, as well as the description of a
generalization of that model. This task will allow us to argue, in Section
VIII, that decoherence is relative to the decomposition of the whole closed
system into a decoherent system and its environment. On this basis, in Section
IX we will cast a new look at the phenomenon of decoherence by considering how
the new theoretical framework dissolves the conceptual difficulties of the EID
approach, and by discussing the relationship between decoherence and
dissipation. Section X will be devoted to describe the quantum-to-classical
transition from the new closed-system approach, which will allow us to
unequivocally define the realm of classicality by identifying the observables
with classical-like behavior. In Section XI we will compare our development
with other approaches to decoherence. Finally, in Section XII we will resume
the main results of our paper and draw our conclusions.

\section{The historical development of the decoherence program}

From a historical perspective, the decoherence program finds its origin
$-$though, of course, not under this name$-$ in the attempts to explain how a
coherent pure state becomes a final decohered mixture with no interference
terms. Three general periods can be identified in the development of this program:

\begin{itemize}
\item \textbf{First period (}closed systems). In the fifties and the early
sixties, some authors directed their attention to the emergence of classical
macroscopic features from quantum microscopic descriptions (van Kampen
\cite{van Kampen}, van Hove \cite{van Hove-1}, \cite{van Hove-2}, Daneri
\textit{et al.} \cite{Daneri}). In this period, the issue was treated in the
context of the study of irreversibility and, therefore, closed systems were
considered. On this basis, the states indistinguishable from the viewpoint of
certain \textquotedblleft gross\textquotedblright\ observables were described
by the same coarse-grained state, whose evolution was proved to reach
equilibrium in a certain relaxation time. The main problem of this period was
that the relaxation times so obtained turned out to be too long to account for
experimental results (see \cite{Omnes-2005}).

\item \textbf{Second period} (open systems). In the seventies, the emergence
of classicality begun to be conceived in terms of quantum measurement and,
then, was addressed from an open-system perspective (Zeh \cite{Zeh-1970},
\cite{Zeh-1971}, \cite{Zeh-1973}). On the basis of these precedents, the EID
approach was systematized and developed mainly by Zurek and his collaborators
in a great number of works (\cite{Zurek-1981}, \cite{Zurek-1982},
\cite{Zurek-1991}, \cite{Zurek-1993}, \cite{Paz-Zurek-2002}, \cite{Zurek-2003}%
). In this context, an open system is considered in interaction with its
environment, and the evolution of its reduced state is studied. EID proves
that, in many physical models, the interference terms of the reduced state
rapidly vanish and the system decoheres in an extremely short decoherence
time. This result solves the main problem of the first period; however, the
foundations of the EID program are still threatened by certain conceptual
problems derived from its open-system perspective (we will return on this
point in the next section).

\item \textbf{Third period} (open and closed systems). Although `EID' is still
considered almost as a synonym for `decoherence', in the last times other
approaches have been proposed to face the conceptual difficulties of EID
(Diosi \cite{Diosi-1}, \cite{Diosi-2}, Milburn \cite{Milburn}, Penrose
\cite{Penrose}, Casati and Chirikov \cite{Casati-Chirikov-1},
\cite{Casati-Chirikov-2}, Adler \cite{Adler}). Some of these accounts are
clearly non-dissipative (Bonifacio \textit{et al. }\cite{Bonifacio}, Ford and
O'Connell \cite{Ford}, Frasca \cite{Frasca}, Sicardi Schifino \textit{et al.}
\cite{Sicardi}, Gambini \textit{et al}. \cite{Gambini}), that is, not based on
the dissipation of energy from the system to the environment. Among them, the
self-induced decoherence (SID) approach shows that a closed quantum system
with continuous spectrum may decohere by destructive interference
(\cite{Castagnino-Laura-2000a}, \cite{Castagnino-Laura-2000b},
\cite{Castagnino-Lombardi-2003}, \cite{Castagnino-Lombardi-2004},
\cite{Castagnino-Lombardi-2005a}, \cite{Castagnino-Lombardi-2005b},
\cite{Castagnino-Gadella}, \cite{Castagnino-2006}).
\end{itemize}

In spite of the fact that, at present, formalisms for open and closed systems
coexist, in the literature both kinds of approaches are often presented as
alternative scenarios for decoherence, and even as theories dealing with
different physical phenomena. In the next sections we will challenge this
common view by showing that the different formalisms can be understood in the
context of a general closed-system framework.

\section{The conceptual difficulties of environment-induced decoherence}

In spite of the great success of EID, and of the fact that it is conceived as
the orthodox approach for decoherence, it still has to face two conceptual difficulties.

\subsection{The \textquotedblleft open-system\textquotedblright\ problem}

According to EID, decoherence is a consequence of the interaction between an
open system and its environment; this process is what \textquotedblleft
einselects\textquotedblright\ the quantum states that become the candidates to
classical states. Therefore, decoherence must always be accompanied by other
manifestations of openness, such as the dissipation of energy into the
environment. Zurek even considers that the prejudice which seriously delayed
the solution of the problem of the emergence of classicality is itself rooted
in the fact that the role of the openness of the system was traditionally
ignored (\cite{Paz-Zurek-2002}, \cite{Zurek-2003}).

If only open systems may decohere, from this perspective the issue of the
emergence of classicality in closed systems, in particular, in the Universe as
a whole, cannot even be posed (see \cite{Pessoa}). Zurek expresses the
criticism to EID in the following terms: \textquotedblleft\textit{the Universe
as a whole is still a single entity with no `outside' environment, and,
therefore, any resolution involving its division into systems is
unacceptable}\textquotedblright\ (\cite{Zurek-1994}, p.181). This objection
has led to the development of the non-dissipative approaches to decoherence
which, for this reason, are usually viewed as alternative or rival to the EID approach.

\subsection{The \textquotedblleft defining systems\textquotedblright\ problem}

When the EID approach is applied to cosmology, the Universe is split into some
degrees of freedom representing the system, and the remaining degrees of
freedom that are supposed to be non accessible and, therefore, play the role
of the environment (see, e.g., \cite{Calzetta et al}). The same strategy is
followed in the case of \textquotedblleft internal\textquotedblright%
\ environments, such as collections of phonons or other internal excitations.
The possibility of \textquotedblleft internal\textquotedblright\ environments
shows that EID supplies no general criterion for distinguishing between the
system and its environment: the partition of the whole closed system is
decided case by case, and usually depends on the previous assumption of the
observables that will behave classically (see discussion in
\cite{Castagnino-Lombardi-2004}).

The absence of a general criterion for deciding where to place the
\textquotedblleft cut\textquotedblright\ between system and environment is a
serious difficulty for an approach that insists on the essential role played
by the openness of the system in the emergence of classicality. Zurek
recognizes this problem as a shortcoming of his proposal: \textquotedblleft%
\textit{In particular, one issue which has been often taken for granted is
looming big as a foundation of the whole decoherence program. It is the
question of what are the `systems' which play such a crucial role in all the
discussions of the emergent classicality. This issue was raised earlier, but
the progress to date has been slow at best}\textquotedblright%
\ (\cite{Zurek-1998}, p.22).\bigskip

As we will see, these problems, which seem to be serious conceptual obstacles
for the EID approach, loose their original strength when decoherence is
understood from a new general perspective. For this purpose, in the next two
sections we will begin by comparing the open-system perspective and the
closed-system perspective.

\section{The open-system perspective}

As it is well-known, given a closed system $S_{T}$ represented in the Hilbert
space $\mathcal{H}_{T}$, its state $\rho_{T}(t)\in\mathcal{H}_{T}%
\otimes\mathcal{H}_{T}$ evolves according to the Liouville-von Neumann
equation, $i\hbar\,d\rho_{T}/dt=\left[  H,\rho_{T}\right]  $,\ where $H$ is
the system's Hamiltonian. If the task is to describe the behavior of $S_{T}%
$\ as a whole, there is no theoretical difference between using its state
$\rho_{T}$ and using the expectation values $\left\langle O\right\rangle
_{\rho_{T}}$ of its observables $O\in\mathcal{H}_{T}\otimes\mathcal{H}_{T}$:
given the state, we can compute the expectation values of all the observables
of the system, and given the expectation values of all the observables (or of
the projectors corresponding to a basis of the Hilbert space), we can compute
the system's state. Therefore, in this case both descriptions are equivalent:
although the state-based description is the commonly used one, all the
information of physical interest is also given by the expectation values of
the system's observables.

If the system is composite, its initial state can be obtained as the tensor
product of the states of its subsystems. For example, in the case of two
subsystems $S_{1}$ and $S_{2}$, represented in the Hilbert spaces
$\mathcal{H}_{1}$ and $\mathcal{H}_{2}$ such that $\mathcal{H}_{T}=$
$\mathcal{H}_{1}\otimes\mathcal{H}_{2}$ , and in initial states $\rho_{1}(0)$
and $\rho_{2}(0)$ respectively, the initial state of the composite system
$S_{T}=S_{1}\cup S_{2}$ is computed as $\rho_{T}(0)=\rho_{1}(0)\otimes\rho
_{2}(0)$ and, conversely, the initial states of the subsystems can be obtained
from the initial state of the composite system by means of the operation of
partial trace:%
\begin{equation}
\rho_{1}(0)=Tr_{2}\left(  \rho_{T}(0)\right)  \in\mathcal{H}_{1}%
\otimes\mathcal{H}_{1}\text{ \ \ \ and \ \ \ }\rho_{2}(0)=Tr_{1}\left(
\rho_{T}(0)\right)  \in\mathcal{H}_{2}\otimes\mathcal{H}_{2} \label{1}%
\end{equation}

The question is how to describe the behavior of the subsystems \textit{at any
time}.

\subsection{States and observables from the open-system perspective}

The usual strategy consists in generalizing the partial trace procedure for
all times by considering that the reduced states, obtained as%
\begin{equation}
\rho_{1}(t)=Tr_{2}\left(  \rho_{T}(t)\right)  \in\mathcal{H}_{1}%
\otimes\mathcal{H}_{1}\text{ \ \ \ and \ \ \ }\rho_{2}(t)=Tr_{1}\left(
\rho_{T}(t)\right)  \in\mathcal{H}_{2}\otimes\mathcal{H}_{2} \label{2}%
\end{equation}
are the quantum states of the subsystems $S_{1}$ and $S_{2}$ in the same sense
as $\rho_{T}(t)$ is the quantum state of the whole closed system $S_{T}$ (a
strategy that is at least controversial: see \cite{dEspagnat-1976},
\cite{dEspagnat-1995}, \cite{Schloss-2007}; we will come back to this point
below). However, unlike $\rho_{T}(t)$, the reduced states $\rho_{1}(t)$ and
$\rho_{2}(t)$ do not always evolve according to the unitary Liouville-von
Neumann equation: in the case of interacting subsystems, they evolve according
to non-unitary master equations, whose specific forms depend on the particular
features of the involved subsystems and, as a consequence, must be constructed
case by case.

Nevertheless, if $O_{1}\in\mathcal{H}_{1}\otimes\mathcal{H}_{1}$ is any
observable of $S_{1}$ and $O_{2}\in\mathcal{H}_{2}\otimes\mathcal{H}_{2}$ is
any observable of $S_{2}$, the expectation values of these observables can be
computed as%
\begin{equation}
\left\langle O_{1}\right\rangle _{\rho_{1}(t)}=Tr\left(  \rho_{1}%
(t)O_{1}\right)  \text{ \ \ \ and \ \ \ }\left\langle O_{2}\right\rangle
_{\rho_{2}(t)}=Tr\left(  \rho_{2}(t)O_{2}\right)  \label{3}%
\end{equation}
In fact, the reduced state of a subsystem of a larger system is defined
precisely as the density operator by means of which the expectation values of
all the observables belonging to that subsystem can be computed. This means
that if we focus only on, say, $S_{1}$, we can rely on the description given
by its reduced state $\rho_{1}(t)$, since it supplies all the information that
we can obtain when we have experimental access only to that subsystem.
Therefore, analogously to the closed-system case, in this situation it is also
true that there is no difference between describing the open system $S_{1}$ by
means of its reduced state $\rho_{1}(t)$ and describing it by means of the
expectation values $\left\langle O_{1}\right\rangle _{\rho_{1}(t)}$ of its observables.

\subsection{Environment-induced decoherence: an open-system approach}

The orthodox EID approach considers the system $S$ under study $-$represented
in a Hilbert space $\mathcal{H}_{S}-$ in interaction with an environment $E$
$-$represented in a Hilbert space $\mathcal{H}_{E}-$ that induces decoherence.
This approach is based on the analysis of the evolution of the reduced state
$\rho_{S}(t)$ of $S$ represented in a certain \textquotedblleft
pointer\textquotedblright\ basis of $\mathcal{H}_{S}$. Either by explicitly
computing $\rho_{S}(t)$ or by analyzing the master equation, it can be
determined whether, under certain conditions, the reduced state operator
becomes diagonal or not in that basis. The diagonalization of its reduced
state is viewed as a manifestation of the decoherence of the open system $S$.

The EID approach proves that, in many physical models, the non-diagonal terms
of the reduced state rapidly tend to vanish after an extremely short
decoherence time $t_{D}$: \qquad\qquad\qquad\qquad\qquad\qquad\qquad\qquad%
\begin{equation}
\rho_{S}(t)\overset{t\gg t_{D}}{\longrightarrow}\rho_{S}^{d} \label{4}%
\end{equation}
where $\rho_{S}^{d}$ is diagonal in the pointer basis. Thus, it is usually
said that the system $S$ decoheres as a consequence of its interaction with
the large number of degrees of freedom of the environment $E$.

We know that the state $\rho_{S}(t)$ is a Hermitian operator, so it can always
be diagonalized in the Smith basis. So, what is special about the state
diagonalization in EID? According to EID, the diagonalization of the state
must be studied in the \textquotedblleft preferred basis\textquotedblright\ or
\textquotedblleft pointer basis\textquotedblright. Pointer states have some
very special proprieties namely \cite{Zurek2009}:

\begin{enumerate}
\item[1.] They are single out by the dynamics, remaining basically unchanged and

\item They result in the smallest entropy increase.
\end{enumerate}

\section{The closed-system perspective}

\subsection{States and observables from the closed-system perspective}

Although the open-system perspective is what underlies the usual practice in
physics, it is not unavoidable. The behavior of the open subsystems can also
be described from a closed-system perspective by considering the expectation
values of the observables relevant in any case. For example, if we still focus
only on the subsystem $S_{1}$, the relevant observables $O_{R}\in
\mathcal{H}_{T}\otimes\mathcal{H}_{T}$ of the whole closed system $S_{T}%
=S_{1}\cup S_{2}$ are those that act only on that subsystem:%
\begin{equation}
O_{R}=O_{1}\otimes I_{2}\in\mathcal{H}_{T}\otimes\mathcal{H}_{T} \label{5}%
\end{equation}
where the $O_{1}\in\mathcal{H}_{1}\otimes\mathcal{H}_{1}$ are the observables
of $S_{1}$, and $I_{2}\in\mathcal{H}_{2}\otimes\mathcal{H}_{2}$ is the
identity of the space of observables of $S_{2}$. The expectation values of
these relevant observables can be computed as%
\begin{equation}
\left\langle O_{R}\right\rangle _{\rho_{T}(t)}=Tr\left(  \rho_{T}%
(t)O_{R}\right)  =Tr\left(  \rho_{T}(t)\left(  O_{1}\otimes I_{2}\right)
\right)  =Tr\left(  \rho_{1}(t)O_{1}\right)  =\left\langle O_{1}\right\rangle
_{\rho_{1}(t)} \label{6}%
\end{equation}
\qquad Eq.(\ref{6}) clearly shows that, if we want to describe only $S_{1}$,
its reduced state is not indispensable. The physically relevant information
about that subsystem can also be obtained by studying the state $\rho_{T}(t)$
of the whole closed system $S_{T}$ and its relevant observables $O_{R}%
=O_{1}\otimes I_{2}$. This means that there is no difference between
describing the open system $S_{1}$ by means of its reduced state $\rho_{1}(t)$
and describing it from a closed-system perspective by means of the expectation
values of the relevant observables $O_{R}=O_{1}\otimes I_{2}$ of the closed
composite system $S_{T}$ in the state $\rho_{T}(t)$.

\subsection{Environment-induced decoherence from a closed-system perspective}

Although the EID approach relies on the diagonalization of the reduced state,
the same phenomenon can also be described from a closed-system perspective (we
will consider the discrete case, but analogous arguments can be developed in
the continuous case). In fact, analogously to the case of the previous
subsection, if we call the whole closed system $U=S\cup E$, represented in the
Hilbert space $\mathcal{H}_{U}$ and whose state is $\rho_{U}(t)$, in this case
the relevant observables of $U$ are of the form%
\begin{equation}
O_{R}=O_{S}\otimes I_{E}\in\mathcal{H}_{U}\otimes\mathcal{H}_{U} \label{7}%
\end{equation}
where the $O_{S}\in\mathcal{H}_{S}\otimes\mathcal{H}_{S}$ are the observables
of $S$, and $I_{E}\in\mathcal{H}_{E}\otimes\mathcal{H}_{E}$ is the identity of
the space of observables of $E$. The expectation values of these relevant
observables can be computed as (see eq.(\ref{6}))%
\begin{equation}
\left\langle O_{R}\right\rangle _{\rho_{U}(t)}=Tr\left(  \rho_{U}%
(t)O_{R}\right)  =Tr\left(  \rho_{U}(t)\left(  O_{S}\otimes I_{E}\right)
\right)  =Tr\left(  \rho_{S}(t)O_{S}\right)  =\left\langle O_{S}\right\rangle
_{\rho_{S}(t)} \label{8}%
\end{equation}

It is clear that the evolution of the reduced state $\rho_{S}(t)$ (see
eq.(\ref{4})) has its counterpart in the evolution of the expectation values:%

\begin{equation}
\left\langle O_{S}\right\rangle _{\rho_{S}(t)}\overset{t\gg t_{D}%
}{\longrightarrow}\left\langle O_{S}\right\rangle _{\rho_{S}^{d}}=\sum_{i}%
\rho_{S{}ii}^{d}\,O_{S{}ii} \label{9}%
\end{equation}
where the $\rho_{S{}ii}^{d}\,$\ and the $O_{S{}ii}$ are the diagonal
components of $\rho_{S}^{d}$ and of $O_{S}$ in the pointer basis,
respectively. But since $\left\langle O_{R}\right\rangle _{\rho_{U}%
(t)}=\left\langle O_{S}\right\rangle _{\rho_{S}(t)}$ (see eq.(\ref{8})), from
the closed-system perspective we can study the time-evolution of the
expectation value $\left\langle O_{R}\right\rangle _{\rho_{U}(t)}$: if, for
any $O_{R}$, such a function tends to settle down, in an extremely short time,
in a value $k=\sum_{i}P_{i}\,O_{S{}ii}$, where $0\leq P_{i}\leq1$ and
$\sum_{i}P_{i}$, then it can be said that $P_{i}\,=\rho_{S{}ii}^{d}$ (see
eq.(\ref{9})). As a consequence,%
\begin{equation}
\left\langle O_{R}\right\rangle _{\rho_{U}(t)}=\left\langle O_{S}\right\rangle
_{\rho_{S}(t)}\overset{t\gg t_{D}}{\longrightarrow}\sum_{i}P_{i}\,O_{S{}%
ii}=\sum_{i}\rho_{S{}ii}^{d}\,O_{S{}ii}=\left\langle O_{S}\right\rangle
_{\rho_{S}^{d}} \label{10}%
\end{equation}
and one can also say that the system has decohered in the pointer basis (see
\cite{Castagnino-Laura-Lombardi-2007}). Moreover, any expectation value
$\left\langle O_{S}\right\rangle _{\rho_{S}^{d}}$ can also be expressed as the
expectation value of the corresponding relevant observable $O_{R}=O_{S}\otimes
I_{E}$ in a coarse-grained state $\rho_{U}^{cg}\in\mathcal{H}_{U}%
\otimes\mathcal{H}_{U}$:%
\begin{equation}
\left\langle O_{S}\right\rangle _{\rho_{S}^{d}}=\left\langle O_{R}%
\right\rangle _{\rho_{U}^{cg}} \label{11}%
\end{equation}
The state $\rho_{U}^{cg}$ can be obtained as $\rho_{S}^{d}\otimes
\widetilde{\delta}_{E}$, where $\widetilde{\delta}_{E}\in\mathcal{H}%
_{E}\otimes\mathcal{H}_{E}$ is a normalized identity operator with
coefficients $\widetilde{\delta}_{E\alpha\beta}=\delta_{\alpha\beta}/%
%TCIMACRO{\dsum \nolimits_{\gamma}}%
%BeginExpansion
{\displaystyle\sum\nolimits_{\gamma}}
%EndExpansion
\delta_{\gamma\gamma}$. It is quite clear that $\rho_{U}^{cg}$ is not the
quantum state of the closed system $U$; nevertheless, if we trace off the
degrees of freedom of the environment, we recover the diagonalized reduced
state of the system:
\begin{equation}
Tr_{E}(\rho_{U}^{cg})=\rho_{S}^{d} \label{12}%
\end{equation}
This means that the coarse-grained state $\rho_{U}^{cg}$ supplies the same
information about the open system $S$ as the reduced state $\rho_{S}^{d}$, but
now from the viewpoint of the closed system $U$. Therefore, from eq.(\ref{10})
we can conclude that%
\begin{equation}
\left\langle O_{R}\right\rangle _{\rho_{U}(t)}\overset{t\gg t_{D}%
}{\longrightarrow}\sum_{i}P_{i}\,O_{S{}ii}=\left\langle O_{R}\right\rangle
_{\rho_{U}^{cg}} \label{13}%
\end{equation}

By following the EID approach, up to this point we have analyzed the case
based on a particular set of relevant observables. In the next section we will
see that the conception of the phenomenon of decoherence can be generalized
when different sets of relevant observables are considered.

\section{Decoherence: a general closed-system approach}

As emphasized by Omn\'{e}s \cite{Omnes1}, decoherence is just a particular
case of the general problem of irreversibility in quantum mechanics. The
problem of irreversibility can be roughly expressed in the following terms.
Since the quantum state $\rho(t)$ follows an unitary evolution, it cannot
reach a final equilibrium state for $t\rightarrow\infty$. Therefore, if the
non-unitary evolution towards equilibrium is to be accounted for, a further
element has to be added to the unitary evolution. From the most general
viewpoint, this element consists in the splitting of the maximal information
about the system into a relevant part and an irrelevant part: whereas the
irrelevant part is disregarded, the relevant part is retained and its
evolution may reach a final equilibrium situation.

This broadly expressed idea can be rephrased in operators language. The
maximal information about the system is given by the set of all its
potentially possible observables. By selecting a particular subset
$\mathcal{O}_{R}$ of this set, we restrict the maximal information to a
relevant part: the expectation values $\langle O_{R}\rangle_{\rho(t)}$ of the
observables $O_{R}\in\mathcal{O}_{R}$ express the relevant information about
the system. Of course, the decision about which observables are to be
considered as relevant depends on the particular purposes in each situation;
but without this restriction, irreversible evolutions cannot be described. As
we have said above, when EID is described from the closed-system perspective,
the phenomenon can be viewed as a particular kind of time-evolution of the
expectation values of certain relevant observables of the closed system $U$.
In that particular case, the relevant observables are of the form $O_{R}%
=O_{S}\otimes I_{E}$. The question is whether that account of decoherence can
be generalized (see \cite{General Framework} for detalis).

Let us begin by considering the expectation values of certain relevant
observables $O_{R}\in\mathcal{H}_{U}\otimes\mathcal{H}_{U}$, which now are not
necessarily of the form $O_{R}=O_{S}\otimes I_{E}$. Let us suppose that, for
any $O_{R}$, those expectation values tend to settle down, in an extremely
short time, in a value $k=\sum_{i}P_{i}\,O_{R{}ii}$, where $0\leq P_{i}\leq1$
and $\sum_{i}P_{i}$ (see the analogous case in EID in eq.(\ref{10})):%
\begin{equation}
\left\langle O_{R}\right\rangle _{\rho_{U}(t)}\overset{t\gg t_{D}%
}{\longrightarrow}\sum_{i}P_{i}\,O_{R{}ii} \label{14}%
\end{equation}
Of course, the $P_{i}$ are not the diagonal elements of a time-independent
$\rho_{U}$, since the state $\rho_{U}(t)$ of the closed system evolves
unitarily. Nevertheless, the sum of eq.(\ref{14}) can also be expressed as
\begin{equation}
\sum_{i}P_{i}\,O_{R{}ii}=\sum_{i}\rho_{Uii}^{d}\,O_{R{}ii} \label{15}%
\end{equation}
where the $\rho_{Uii}^{d}$ can be conceived as the components of a kind of
coarse-grained state $\rho_{U}^{d}$, diagonal in a basis $\left\{  \left\vert
\alpha_{i}\right\rangle \right\}  $ of $\mathcal{H}_{U}$, which plays the role
of pointer basis. In other words, $\left\langle O_{R}\right\rangle _{\rho
_{U}(t)}$ converges, in an extremely short time, to a value that can be
computed as if the system were in a state $\rho_{U}^{d}$ represented by a
diagonal density operator :
\begin{equation}
\left\langle O_{R}\right\rangle _{\rho_{U}(t)}\overset{t\gg t_{D}%
}{\longrightarrow}\sum_{i}P_{i}\,O_{R{}ii}=\sum_{i}\rho_{Uii}^{d}\,O_{R{}%
ii}=\left\langle O_{R}\right\rangle _{\rho_{U}^{d}} \label{16}%
\end{equation}

If eq.(\ref{13}) and eq.(\ref{16}) are compared, it is easy to see that in
both cases the expectation values of certain relevant observables of the whole
closed system tend very rapidly to certain time-independent values that can be
computed as expectation values of those relevant observables in a
time-independent diagonal state. Therefore, if eq.(\ref{13}) describes
decoherence, there is no reason to deny that eq.(\ref{16}) is also a
description of the same phenomenon. This means that, although the off-diagonal
terms of $\rho_{U}(t)$ never vanish through the unitary evolution, decoherence
obtains because it is a coarse-grained process: the system decoheres from the
observational viewpoint given by any observable belonging to the space
$\mathcal{O}_{R}$.

It might be objected that, whereas in the case of EID the time-independent
diagonal state $\rho_{U}^{cg}$ is defined in terms of the reduced state
$\rho_{S}^{d}$, which is the quantum state of the open system, in the general
approach the time-independent diagonal $\rho_{U}^{d}$ is not the quantum state
of the closed system. However, this objection is based on the assumption that
the reduced state of an open system is its quantum state, a position that is
seriously challenged by the difference between proper and improper mixtures
(see \cite{dEspagnat-1976}, \cite{dEspagnat-1995}): reduced states are
improper mixtures, they are only \textquotedblleft\textit{a calculational
tool}\textquotedblright\ for computing expectation values (\cite{Schloss-2007}%
, p.48). For this reason, even in the particular context of decoherence,
Schlosshauer warns us \textquotedblleft\textit{against a misinterpretation of
reduced density matrices as describing a proper mixture of states}%
\textquotedblright\ (\cite{Schloss-2007}, p.69; see also \cite{Zeh-2005}).
Therefore, although it is true that $\rho_{U}^{d}$ is not the state of the
closed system $U$ but a calculational tool for computing expectation values
after the extremely short decoherence time, the same can be said of the
reduced state $\rho_{S}^{d}$ regarding the open system $S$.

It is interesting to emphasize in what sense this closed-system perspective
for understanding decoherence is more general than the orthodox open-system
perspective. The EID approach always selects a set of relevant observables
that correspond to a subsystem, with its associated Hilbert space and its
reduced state. From the closed-system perspective, by contrast, the set of the
relevant observables that decohere is completely generic and may not
correspond to a subsystem. For example, one might be interested in the
decoherence of a single observable of the closed system $U$, which certainly
does not define a subsystem but, nevertheless, may manifest a classical-like behavior.

Summing up, the closed-system perspective has allowed us to conceptualize
decoherence from a viewpoint more general than that supplied by the EID
approach: decoherence \textquotedblleft induced\textquotedblright\ by the
environment is a particular case of a phenomenon that involves the evolution
of the expectation values of certain relevant observables towards a specific
time-independent value. There are, then, certain processes that fall under the
concept of decoherence from this general viewpoint, but are not conceptualized
as decoherence from the EID approach. It might be thought that this is a
merely semantic issue; however, the fact that the difference has physical
content will become clear when the problem of the transition from quantum to
classical will be considered.

\section{The spin-bath model from the closed-system approach}

In this section we will analyze the spin-bath model, both in the traditional
and in the generalized versions, as presented in previous works, from the
open-system approach. We will show how this model leads to conclusions that
sound paradoxical when read from an open-system perspective.

\subsection{The traditional spin-bath model}

This is a very simple model that has been exactly solved in previous papers
(\cite{Zurek-1982}). Here we will study it from the closed-system perspective
presented in the previous section.

Let us consider a closed system $U=P\cup P_{1}\cup P_{2}\cup...\cup
P_{N}=P\cup\left(
%TCIMACRO{\tbigcup \nolimits_{i=1}^{N}}%
%BeginExpansion
{\textstyle\bigcup\nolimits_{i=1}^{N}}
%EndExpansion
P_{i}\right)  $, where (i) $P$ is a spin-1/2 particle represented in the
Hilbert space $\mathcal{H}_{P}$, and (ii) each $P_{i}$ is a spin-1/2 particle
represented in its Hilbert space $\mathcal{H}_{i}$. The Hilbert space of the
composite system is, then,%
\begin{equation}
\mathcal{H}=\left(  \mathcal{H}_{P}\right)  \otimes\left(  \bigotimes
\limits_{i=1}^{N}\mathcal{H}_{i}\right)  \label{17}%
\end{equation}
In the particle $P$, the two eigenstates of the spin operator
$S_{S,\overrightarrow{v}}$\ in direction $\overrightarrow{v}$ are $\left\vert
\Uparrow\right\rangle $ and $\left\vert \Downarrow\right\rangle $, such that
$S_{S,\overrightarrow{v}}\left\vert \Uparrow\right\rangle =\frac{1}%
{2}\left\vert \Uparrow\right\rangle $\ and\ $S_{S,\overrightarrow{v}%
}\left\vert \Downarrow\right\rangle =-\frac{1}{2}\left\vert \Downarrow
\right\rangle $. In each particle $P_{i}$, the two eigenstates of the
corresponding spin operator $S_{i,\overrightarrow{v}}$\ in direction
$\overrightarrow{v}$ are $\left\vert \uparrow_{i}\right\rangle $ and
$\left\vert \downarrow_{i}\right\rangle $, such that $S_{i,\overrightarrow{v}%
}\left\vert \uparrow_{i}\right\rangle =\frac{1}{2}\left\vert \uparrow
_{i}\right\rangle $ and\ $S_{i,\overrightarrow{v}}\left\vert \downarrow
_{i}\right\rangle =\frac{1}{2}\left\vert \downarrow_{i}\right\rangle $.
Therefore, a pure initial state of $U$ reads%
\begin{equation}
|\psi_{0}\rangle=(a\left\vert \Uparrow\right\rangle +b\left\vert
\Downarrow\right\rangle )\bigotimes_{i=1}^{N}(\alpha_{i}|\uparrow_{i}%
\rangle+\beta_{i}|\downarrow_{i}\rangle) \label{18}%
\end{equation}
where the coefficients $a$, $b$, $\alpha_{i}$, $\beta_{i}$ are such that
satisfy $\left\vert a\right\vert ^{2}+\left\vert b\right\vert ^{2}=1$ and
$\left\vert \alpha_{i}\right\vert ^{2}+\left\vert \beta_{i}\right\vert ^{2}%
=1$. If the self-Hamiltonians $H_{P}$ of $P$ and $H_{i}$ of $P_{i}$ are taken
to be zero, and there is no interaction among the $P_{i}$, then the total
Hamiltonian $H$ of the composite system is given by the interaction between
the particle $P$ and each particle $P_{i}$. For instance (see
\cite{Zurek-1982}),%

\begin{equation}
H=\frac{1}{2}\left(  \left\vert \Uparrow\right\rangle \left\langle
\Uparrow\right\vert -\left\vert \Downarrow\right\rangle \left\langle
\Downarrow\right\vert \right)  \otimes\sum_{i=1}^{N}g_{i}\left(  \left\vert
\uparrow\right\rangle \left\langle \uparrow\right\vert -\left\vert
\downarrow\right\rangle \left\langle \downarrow\right\vert \right)
\otimes\left(  \bigotimes_{\substack{j=1 \\j\neq i}}^{N}I_{j}\right)
\label{19}%
\end{equation}
where $I_{j}=\left\vert \uparrow_{j}\right\rangle \left\langle \uparrow
_{j}\right\vert +\left\vert \downarrow_{j}\right\rangle \left\langle
\downarrow_{j}\right\vert $ is the identity operator on the subspace
$\mathcal{H}_{j}$ and the $g_{i}$ are the coupling constants.

\subsubsection{Decomposition 1}

In the typical situation studied by the EID approach, the open system $S$ is
the particle $P$ and the remaining particles $P_{i}$ play the role of the
environment $E$: $S=P$ and $E=%
%TCIMACRO{\tbigcup \nolimits_{i=1}^{N}}%
%BeginExpansion
{\textstyle\bigcup\nolimits_{i=1}^{N}}
%EndExpansion
P_{i}$. Then, the decomposition for this case is%
\begin{equation}
\mathcal{H}=\mathcal{H}_{S}\otimes\mathcal{H}_{E}=\left(  \mathcal{H}%
_{P}\right)  \otimes\left(  \bigotimes\limits_{i=1}^{N}\mathcal{H}_{i}\right)
\label{20}%
\end{equation}
and the relevant observables $O_{R}$ of $U$ are those corresponding to the
particle $P$:%
\begin{equation}
O_{R}=O_{S}\otimes I_{E}=\left(  s_{\Uparrow\Uparrow}\left\vert \Uparrow
\right\rangle \left\langle \Uparrow\right\vert +s_{\Uparrow\Downarrow
}\left\vert \Uparrow\right\rangle \left\langle \Downarrow\right\vert
+s_{\Downarrow\Uparrow}\left\vert \Downarrow\right\rangle \left\langle
\Uparrow\right\vert +s_{\Downarrow\Downarrow}\left\vert \Downarrow
\right\rangle \left\langle \Downarrow\right\vert \right)  \otimes\left(
\bigotimes_{i=1}^{N}I_{i}\right)  \label{21}%
\end{equation}
The expectation value of these observables in the state $|\psi(t)\rangle
=|\psi_{0}\rangle\,e^{-iHt}$ is given by (see
\cite{Castagnino-Fortin-Lombardi-2010a})%
\begin{equation}
\langle O_{R}\rangle_{\psi(t)}=|a|^{2}\,s_{\Uparrow\Uparrow}+|b|^{2}%
\,s_{\Downarrow\Downarrow}+2\operatorname{Re}[ab^{\ast}\,s_{\Downarrow
\Uparrow}\,r(t)]=%
%TCIMACRO{\tsum \nolimits^{d}}%
%BeginExpansion
{\textstyle\sum\nolimits^{d}}
%EndExpansion
+%
%TCIMACRO{\tsum \nolimits^{nd}}%
%BeginExpansion
{\textstyle\sum\nolimits^{nd}}
%EndExpansion
(t) \label{22}%
\end{equation}
where%
\begin{equation}
r(t)=\left\langle \varepsilon_{\Downarrow}(t)\right\vert \varepsilon
_{\Uparrow}(t)\rangle=\prod_{i=1}^{N}\left[  |\alpha_{i}|^{2}e^{ig_{i}%
t}+|\beta_{i}|^{2}e^{-ig_{i}t}\right]  \label{23}%
\end{equation}
By means of numerical simulations it is shown that, for $N\gg1$, in general
$\left\vert r(t)\right\vert ^{2}\rightarrow0$ and, therefore, $%
%TCIMACRO{\tsum \nolimits^{nd}}%
%BeginExpansion
{\textstyle\sum\nolimits^{nd}}
%EndExpansion
(t)\rightarrow0$: the particle $P$ decoheres in interaction with a large
environment $E$ composed by $N$ particles $P_{i}$ (see \cite{Schloss-2007};
for larger values of $N$ and realistic values of the $g_{i}$ in typical models
of spin interaction, see \cite{Castagnino-Fortin-Lombardi-2010a}).

\subsubsection{Decomposition 2}

Although in the usual presentations of the model the system of interest is
$P$, there are different ways of splitting the whole closed system $U$. For
instance, we can decide to observe a particular particle $P_{j}$ of what was
previously considered the environment, and to consider the remaining particles
as the new environment: $S=P_{j}$ and $E=P\cup\left(
%TCIMACRO{\tbigcup \nolimits_{i=1,i\neq j}^{N}}%
%BeginExpansion
{\textstyle\bigcup\nolimits_{i=1,i\neq j}^{N}}
%EndExpansion
P_{i}\right)  $. The total Hilbert space of the closed composite system $U$ is
still given by eq.(\ref{17}), but now the decomposition is%
\begin{equation}
\mathcal{H}=\mathcal{H}_{S}\otimes\mathcal{H}_{E}=\left(  \mathcal{H}%
_{j}\right)  \otimes\left(  \mathcal{H}_{P}\otimes\left(  \bigotimes
\limits_{\substack{i=1\\i\neq j}}^{N}\mathcal{H}_{i}\right)  \right)
\label{24}%
\end{equation}
and the relevant observables $O_{R}$ of $U$ are those corresponding to the
particle $P_{j}$:%
\begin{equation}
O_{R}=O_{S}\otimes I_{E}=\left(  \xi_{\uparrow\uparrow}^{j}\,|\uparrow
_{j}\rangle\langle\uparrow_{j}|+\xi_{\uparrow\downarrow}^{j}\,|\uparrow
_{j}\rangle\langle\downarrow_{j}|+\xi_{\downarrow\uparrow}^{j}\,|\downarrow
_{j}\rangle\langle\uparrow_{j}|+\xi_{\downarrow\downarrow}^{j}\,|\downarrow
_{j}\rangle\langle\downarrow_{j}|\right)  \otimes\left(  I_{P}\otimes\left(
\bigotimes\limits_{\substack{i=1\\i\neq j}}^{N}I_{i}\right)  \right)
\label{25}%
\end{equation}
The expectation value of these observables in the state $|\psi(t)\rangle$ is
given by (\cite{Castagnino-Fortin-Lombardi-2010a})%
\begin{equation}
\langle O_{R}\rangle_{\psi(t)}=|\alpha_{j}|^{2}\,\xi_{\uparrow\uparrow}%
^{j}+|\beta_{j}|^{2}\xi_{\downarrow\downarrow}^{j}+2\operatorname{Re}%
[\alpha_{j}\beta_{j}^{\ast}\,\xi_{\downarrow\uparrow}^{j}\,e^{ig_{i}t}]=%
%TCIMACRO{\tsum \nolimits^{d}}%
%BeginExpansion
{\textstyle\sum\nolimits^{d}}
%EndExpansion
+%
%TCIMACRO{\tsum \nolimits^{nd}}%
%BeginExpansion
{\textstyle\sum\nolimits^{nd}}
%EndExpansion
(t) \label{26}%
\end{equation}
In this case, numerical simulations are not necessary to see that the
time-dependent term of eq.(\ref{26}) is an oscillating function which,
therefore, has no limit for $t\rightarrow\infty$. This result is not
surprising, but completely reasonable from a physical point of view. In fact,
with the exception of the particle $P$, the remaining particles of the
environment $E$ are uncoupled to each other: each $P_{i}$ evolves as a free
system and, as a consequence, $E$ is unable to reach a final stable state.

\subsection{A generalized spin-bath model}

Let us now consider a closed system $U=A\cup B$ where

\begin{enumerate}
\item[(i)] The subsystem $A$ is composed of $M$ spin-1/2 particles $A_{i}$,
with $i=1,2,...,M$, each one represented in its Hilbert space $\mathcal{H}%
_{A_{i}}$: in each $A_{i}$, the two eigenstates of the spin operator
$S_{A_{i},\overrightarrow{v}}$\ in direction $\overrightarrow{v}$ are
$\left\vert \Uparrow_{i}\right\rangle $ and $\left\vert \Downarrow
_{i}\right\rangle $.

\item[(ii)] The subsystem $B$ is composed of $N$ spin-1/2 particles $B_{k}$,
with $k=1,2,...,N$, each one represented in its Hilbert space $\mathcal{H}%
_{B_{k}}$: in each $B_{k}$, the two eigenstates of the spin operator
$S_{B_{k},\overrightarrow{v}}$\ in direction $\overrightarrow{v}$ are
$\left\vert \uparrow_{k}\right\rangle $ and $\left\vert \downarrow
_{k}\right\rangle $.
\end{enumerate}

The Hilbert space of the composite system $U=A\cup B$ is, then%
\begin{equation}
\mathcal{H}=\mathcal{H}_{A}\otimes\mathcal{H}_{B}=\left(  \bigotimes
\limits_{i=1}^{M}\mathcal{H}_{A_{i}}\right)  \otimes\left(  \bigotimes
\limits_{k=1}^{N}\mathcal{H}_{B_{k}}\right)  \label{27}%
\end{equation}
and a pure initial state of $U$ reads%
\begin{equation}
|\psi_{0}\rangle=|\psi_{A}\rangle\otimes|\psi_{B}\rangle=\left(
\bigotimes\limits_{i=1}^{M}(a_{i}\left\vert \Uparrow_{i}\right\rangle
+b_{i}\left\vert \Downarrow_{i}\right\rangle )\right)  \otimes\left(
\bigotimes\limits_{k=1}^{N}(\alpha_{k}|\uparrow_{k}\rangle+\beta
_{k}|\downarrow_{k}\rangle)\right)  \label{28}%
\end{equation}
with $\left\vert a_{i}\right\vert ^{2}+\left\vert b_{i}\right\vert ^{2}=1$ and
$\left\vert \alpha_{k}\right\vert ^{2}+\left\vert \beta_{k}\right\vert ^{2}%
=1$. As in the original spin-bath model, the self-Hamiltonians $H_{A_{i}}$ and
$H_{B_{k}}$ are taken to be zero, and there is no interaction among the
particles $A_{i}$ nor among the particles $B_{k}$. As a consequence, the total
Hamiltonian of the composite system is given by
\begin{equation}
H=\frac{1}{2}\left(  \left\vert \Uparrow\right\rangle \left\langle
\Uparrow\right\vert -\left\vert \Downarrow\right\rangle \left\langle
\Downarrow\right\vert \right)  \left(  \sum_{i=1}^{M}\frac{1}{2}\left(
\left\vert \Uparrow_{i}\right\rangle \left\langle \Uparrow_{i}\right\vert
-\left\vert \Downarrow_{i}\right\rangle \left\langle \Downarrow_{i}\right\vert
\right)  \otimes\left(  \bigotimes_{\substack{j=1\\j\neq i}}^{M}I_{A_{j}%
}\right)  \right)  \otimes\left(  \sum_{k=1}^{N}g_{k}\left(  \left\vert
\uparrow_{k}\right\rangle \left\langle \uparrow_{k}\right\vert -\left\vert
\downarrow_{k}\right\rangle \left\langle \downarrow_{k}\right\vert \right)
\otimes\left(  \bigotimes_{\substack{l=1\\l\neq k}}^{N}I_{l}\right)  \right)
\label{29}%
\end{equation}
where $I_{A_{j}}=\left\vert \Uparrow_{j}\right\rangle \left\langle
\Uparrow_{j}\right\vert +\left\vert \Downarrow_{j}\right\rangle \left\langle
\Downarrow_{j}\right\vert $ and $I_{B_{k}}=\left\vert \uparrow_{k}%
\right\rangle \left\langle \uparrow_{k}\right\vert +\left\vert \downarrow
_{k}\right\rangle \left\langle \downarrow_{k}\right\vert $ are the identity
operators on the subspaces $\mathcal{H}_{A_{j}}$ and $\mathcal{H}_{B_{k}}$
respectively. Let us notice that the eq.(\ref{19}) of the original model is
the particular case of eq.(\ref{29}) for $M=1$.%

\begin{figure}[t]
%\vspace*{-1.cm}
\par
\centerline{\scalebox{0.7}{\includegraphics{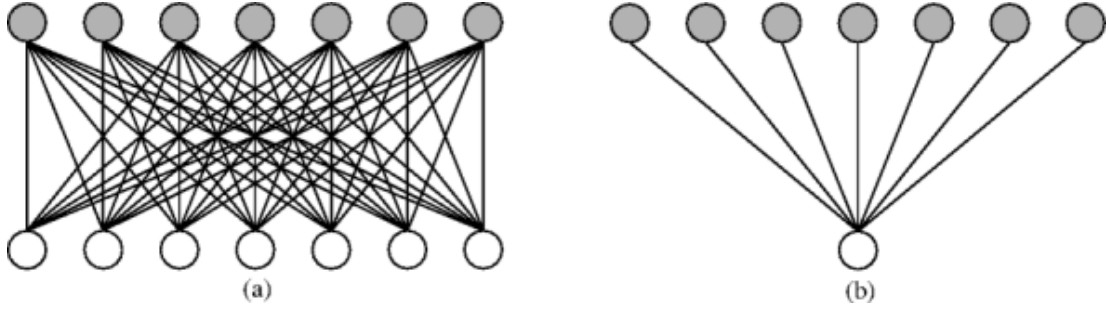}}} 
% \vspace*{-10.25cm}
\vspace*{0.cm}
\caption{Schema of the interactions among the particles: (a) generalized
spin-bath model ($M>1$), and (b) original spin-bath model ($M=1$).}
\label{f1}
\end{figure}

\subsubsection{Decomposition 1}

We can consider the decomposition where $A$ is the open system $S$ and $B$ is
the environment $E$. This is a generalization of Decomposition 1 in the
traditional spin-bath model: the only difference is that here $S$ is composed
of $M\geq1$ particles instead of only one. Then, the decomposition is%
\begin{equation}
\mathcal{H}=\mathcal{H}_{S}\otimes\mathcal{H}_{E}=\left(  \bigotimes
\limits_{i=1}^{M}\mathcal{H}_{A_{i}}\right)  \otimes\left(  \bigotimes
\limits_{k=1}^{N}\mathcal{H}_{B_{k}}\right)  \label{30}%
\end{equation}
and the relevant observables $O_{R}$ are those corresponding to $A$:%

\begin{equation}
O_{R}=O_{S}\otimes I_{E}=O_{A}\otimes\left(  \bigotimes\limits_{i=1}^{N}%
I_{i}\right)  \label{31}%
\end{equation}
When the expectation value $\langle O_{R}\rangle_{\psi(t)}=%
%TCIMACRO{\tsum \nolimits^{d}}%
%BeginExpansion
{\textstyle\sum\nolimits^{d}}
%EndExpansion
+%
%TCIMACRO{\tsum \nolimits^{nd}}%
%BeginExpansion
{\textstyle\sum\nolimits^{nd}}
%EndExpansion
(t)$ of the observables $O_{R}$ in the state $|\psi(t)\rangle$ is computed,
two cases can be distinguished:

\begin{itemize}
\item \textbf{Case (a)}: $M\ll N$

Numerical simulations show that $%
%TCIMACRO{\tsum \nolimits^{nd}}%
%BeginExpansion
{\textstyle\sum\nolimits^{nd}}
%EndExpansion
(t)\rightarrow0$ very fast for increasing time (see Figure 2 of
\cite{Castagnino-Fortin-Lombardi-2010a}). This means that, as expected, a
small open system $S=A$ of $M$\ particles decoheres in interaction with a
large environment $E=B$ of $N\gg M$ particles.

\item \textbf{Case (b)}: $M\gg N$ or $M\simeq N$

Numerical simulations show that $%
%TCIMACRO{\tsum \nolimits^{nd}}%
%BeginExpansion
{\textstyle\sum\nolimits^{nd}}
%EndExpansion
(t)$ exhibits an oscillating behavior and, then, it does not approach zero for
increasing time (see Figures 3 and 4 of
\cite{Castagnino-Fortin-Lombardi-2010a}). This means that, when the
environment $E=B$ of $N$ particles is not large enough when compared with the
open system $S=A$ of $M$\ particles, $S$ does not decohere.
\end{itemize}

\subsubsection{Decomposition 2}

In this case we decide to observe only one particle of $A$. This amounts to
splitting the closed system $U$ into two new subsystems: the open system $S$
is, say, the particle $A_{M}$, and the environment $E$ is $\left(
%TCIMACRO{\tbigcup \nolimits_{i=1}^{M-1}}%
%BeginExpansion
{\textstyle\bigcup\nolimits_{i=1}^{M-1}}
%EndExpansion
A_{i}\right)  \cup\left(
%TCIMACRO{\tbigcup \nolimits_{k=1}^{N}}%
%BeginExpansion
{\textstyle\bigcup\nolimits_{k=1}^{N}}
%EndExpansion
B_{k}\right)  $. Let us notice that the Decomposition 2 of the traditional
spin-bath model is a particular case of this one, for $N=1$ (where $N$ plays
the role of the $M$ of this case). The decomposition here is%
\begin{equation}
\mathcal{H}=\mathcal{H}_{S}\otimes\mathcal{H}_{E}=\left(  \mathcal{H}_{A_{M}%
}\right)  \otimes\left(  \left(  \bigotimes\limits_{i=1}^{M-1}\mathcal{H}%
_{A_{i}}\right)  \otimes\left(  \bigotimes\limits_{k=1}^{N}\mathcal{H}_{B_{k}%
}\right)  \right)  \label{32}%
\end{equation}
and the relevant observables $O_{R}$ of $U$ are those corresponding to $A_{M}%
$:%
\begin{equation}
O_{R}=O_{S}\otimes I_{E}=O_{A_{M}}\otimes\left(  \left(  \bigotimes
\limits_{i=1}^{M-1}I_{i}\right)  \otimes\left(  \bigotimes\limits_{k=1}%
^{N}I_{k}\right)  \right)  \label{33}%
\end{equation}
When the expectation value $\langle O_{R}\rangle_{\psi(t)}=%
%TCIMACRO{\tsum \nolimits^{d}}%
%BeginExpansion
{\textstyle\sum\nolimits^{d}}
%EndExpansion
+%
%TCIMACRO{\tsum \nolimits^{nd}}%
%BeginExpansion
{\textstyle\sum\nolimits^{nd}}
%EndExpansion
(t)$ is computed, numerical simulations show that, if $N\gg1$, $%
%TCIMACRO{\tsum \nolimits^{nd}}%
%BeginExpansion
{\textstyle\sum\nolimits^{nd}}
%EndExpansion
(t)\rightarrow0$ very fast for increasing time (see Figures 5, 6 and 7 of
\cite{Castagnino-Fortin-Lombardi-2010a}). This means that the particle $A_{M}$
decoheres when $N\gg1$, independently of the value of $M$. But since the
particle $A_{M}$ was arbitrarily selected, the same argument holds for any
particle $A_{i}$ of $A$. Then, when $N\gg1$ and independently of the value of
$M$, any particle $A_{i}$ decoheres in interaction with its environment $E$ of
$N+M-1$ particles. On the other hand, the symmetry of the whole system $U$
allows us to draw analogous conclusions when the system $S$ is one of the
particles of $B$: when $M\gg1$ and independently of the value of $N$, any
particle $B_{k}$ decoheres in interaction with its environment $E$ of $N+M-1$ particles.

\subsection{Analyzing results}

Let us consider the generalized spin-bath model when $M=N\gg1$. In this case,
the subsystem $A=%
%TCIMACRO{\tbigcup \nolimits_{i=1}^{M}}%
%BeginExpansion
{\textstyle\bigcup\nolimits_{i=1}^{M}}
%EndExpansion
A_{i}$ does not decohere (Decomposition 1), but the particles $A_{i}$,
considered independently, do decohere (Decomposition 2). In other words, in
spite of the fact that certain particles decohere and may behave classically,
the subsystem composed by all of them retains its quantum nature. We have also
seen that, since $M\gg1$, all the particles $B_{k}$, considered independently,
decohere. Then, in this case not only all the $A_{i}$, but also all the
$B_{k}$ decohere. This means that all the particles of the closed system
$U=\left(
%TCIMACRO{\tbigcup \nolimits_{i=1}^{M}}%
%BeginExpansion
{\textstyle\bigcup\nolimits_{i=1}^{M}}
%EndExpansion
A_{i}\right)  \cup\left(
%TCIMACRO{\tbigcup \nolimits_{k=1}^{N}}%
%BeginExpansion
{\textstyle\bigcup\nolimits_{k=1}^{N}}
%EndExpansion
B_{k}\right)  $ may become classical when considered independently, although
the whole system $U$ certainly does not decohere and, therefore, retains its
quantum character.

The fact that certain particles may be classical or quantum depending on how
they are considered sounds paradoxical in the context of an approach that
explains decoherence as the result of an interaction between open systems.
This difficulty can also be seen as a manifestation of the \textquotedblleft
looming big\textquotedblright\ problem of defining the open systems involved
in decoherence. The irony of this story is that such a problem is the
consequence of what has been considered to be the main advantage of the EID
program, its open-system perspective, according to which particles interacting
with other particles are well-defined open systems, and the collections of
those particles are open systems too. So, the problem is to decide which one
of all these open systems is the system $S$ that decoheres or, in other words,
where to place the cut between the system $S$ and its environment $E$.

The open-system approach not only leads to the \textquotedblleft looming
big\textquotedblright\ problem, but in a certain sense also disregards the
well-known holism of quantum mechanics: a quantum system is not the mere
collection of its parts and the interactions among them. In order to retain
its holistic nature, a quantum system has to be considered as a whole: the
open \textquotedblleft subsystems\textquotedblright\ are only partial
descriptions of the whole closed system. It is on the basis of this
closed-system perspective that a different conceptual viewpoint for
understanding decoherence can be developed.

\section{The Relative Nature of Decoherence}

From the closed-system perspective introduced in the previous sections, the
discrimination between system and environment turns out to be the selection of
the relevant observables. By following Harshman and Wickramasekara
(\cite{Sujeeva}), we will use the expression \textquotedblleft tensor product
structure\textquotedblright\ (TPS) to call any factorization $\mathcal{H}%
=\mathcal{H}_{A}\otimes\mathcal{H}_{B}$ of a Hilbert space $\mathcal{H}$,
defined by the set of observables $\left\{  O_{A}\otimes I_{B},I_{A}\otimes
O_{B}\right\}  $, such that the eigenbases of the sets $\left\{
O_{A}\right\}  $ and $\left\{  O_{B}\right\}  $ are bases of $\mathcal{H}_{A}$
and $\mathcal{H}_{B}$ respectively. If $\mathcal{H}$ corresponds to a closed
system $U$, the TPS $\mathcal{H}=\mathcal{H}_{A}\otimes\mathcal{H}_{B}$
represents the decomposition of $U$ into two open systems $S_{A}$ and $S_{B}$,
corresponding to the Hilbert spaces $\mathcal{H}_{A}$ and $\mathcal{H}_{B}$
respectively. In turn, given the space $\mathcal{O}=\mathcal{H}\otimes
\mathcal{H}$ of the observables of $U$, such a decomposition identifies the
spaces $\mathcal{O}_{A}=\mathcal{H}_{A}\otimes\mathcal{H}_{A}$ and
$\mathcal{O}_{B}=\mathcal{H}_{B}\otimes\mathcal{H}_{B}$ of the observables of
the open systems $S_{A}$ and $S_{B}$, such that $\mathcal{O}_{A}\otimes
I_{B}\subset\mathcal{O}$ and $I_{A}\otimes\mathcal{O}_{A}\subset\mathcal{O}$.
In particular, the total Hamiltonian of $U$, $H\in\mathcal{O}$, can be
expressed as $H=H_{A}\otimes I_{B}+I_{B}\otimes H_{A}+H_{AB}$, where $H_{A}%
\in\mathcal{O}_{A}$ is the Hamiltonian of $S_{A}$, $H_{B}\in\mathcal{O}_{B}$
is the Hamiltonian of $S_{B}$, and $H_{AB}\in\mathcal{O}$ is the interaction
Hamiltonian, representing the interaction between the open systems $S_{A}$ and
$S_{B}$.

In general, a quantum system $U$ admits a variety of TPSs, that is, of
decompositions into $S_{A}$ and $S_{B}$. Among all these possible
decompositions, there may be a particular TPS that remains \textit{dynamically
invariant} (see \cite{Sujeeva}). This is the case when there is no interaction
between $S_{A}$ and $S_{B}$, $H_{AB}=0$, and, then,%
\begin{equation}
\left[  H_{A}\otimes I_{B},I_{B}\otimes H_{A}\right]  =0\quad\Rightarrow
\quad\exp\left(  -iHt\right)  =\exp\left(  -iH_{A}t\right)  \exp\left(
-iH_{B}t\right)  \label{34}%
\end{equation}
Therefore,%
\begin{align}
\rho_{A}(t)  &  =Tr_{B}\rho(t)=Tr_{B}\left(  e^{-iHt}\rho_{0}e^{iHt}\right)
=e^{-iH_{A}t}\left(  Tr_{B}\rho_{0}\right)  e^{iH_{A}t}=e^{-iH_{A}t}\rho
_{A0}e^{iH_{A}t}\label{35}\\
\rho_{B}(t)  &  =Tr_{A}\rho(t)=Tr_{A}\left(  e^{-iHt}\rho_{0}e^{iHt}\right)
=e^{-iH_{B}t}\left(  Tr_{A}\rho_{0}\right)  e^{iH_{B}t}=e^{-iH_{B}t}\rho
_{B0}e^{iH_{B}t} \label{36}%
\end{align}
This means that, even if the initial state $\rho_{0}$ of $U$\ is an entangled
state with respect to the TPS $\mathcal{H}=\mathcal{H}_{A}\otimes
\mathcal{H}_{B}$, $S_{A}$ and $S_{B}$ are \textit{dynamically independent}:
each one evolves unitarily under the action of its own Hamiltonian. As a
consequence, the subsystems $S_{A}$ and $S_{B}$ resulting from this particular
TPS do not decohere.

Once we have excluded the dynamically invariant TPS, all the remaining TPSs of
$U$ define subsystems $S_{A}$ and $S_{B}$ such that $H_{AB}\neq0$. As a result
of the interaction, $S_{A}$ and $S_{B}$ evolve non-unitarily; then, depending
on the particular $H_{AB}$, they may decohere. But the point to stress here is
that there is no privileged non-dynamically invariant decomposition of $U$:
each partition of the closed system into $S_{A}$ and $S_{B}$ is just a way of
selecting the spaces of observables $\mathcal{O}_{A}$ and $\mathcal{O}_{B}$.

Once these concepts are considered, the selection of the space $\mathcal{O}%
_{R}$ of relevant observables in the EID approach amounts to the selection of
a particular TPS, $\mathcal{H}=\mathcal{H}_{S}\otimes\mathcal{H}_{E}$, such
that $\mathcal{O}_{R}=\mathcal{O}_{S}\otimes I_{E}\subset\mathcal{O=H}%
\otimes\mathcal{H}$. From this closed-system perspective, it turns out to be
clear that there is no essential criterion for identifying the
\textquotedblleft open system\textquotedblright\ and its \textquotedblleft
environment\textquotedblright. Given the closed system $U$, that
identification requires two steps: (i) to select a TPS $\mathcal{H}%
=\mathcal{H}_{A}\otimes\mathcal{H}_{B}$ such that $U=S_{A}\cup S_{B}$, and
(ii) to decide that one of the systems resulting from the decomposition, say
$S_{A}$, is the open system $S$, and the other, $S_{B}$, is the environment
$E$. Since the TPS is defined by the spaces of observables $\mathcal{O}_{A}$
and $\mathcal{O}_{B}$, the decomposition of $U$ is just the adoption of a
descriptive perspective: the identification of $S$ and $E$ amounts to the
selection of the relevant observables in each situation. But since the split
can be performed in many ways, with no privileged decomposition, there is no
need of an unequivocal criterion for deciding where to place the cut between
\textquotedblleft the\textquotedblright\ system and \textquotedblleft
the\textquotedblright\ environment. Decoherence is not a yes-or-no process,
but a phenomenon \textit{relative} to the chosen decomposition of the whole
closed quantum system.

From this perspective, the perplexities derived from the generalized spin-bath
model vanish. In fact, when we consider the whole closed system $U$, there is
no difficulty in saying that from the viewpoint of the space of observables,
say $\mathcal{O}_{A_{1}}$, (corresponding to the particle $A_{1}$) there is
decoherence, but from the viewpoint of the space of observables $\mathcal{O}%
_{A}$ (corresponding to the open subsystem $A=%
%TCIMACRO{\tbigcup \nolimits_{i=1}^{M}}%
%BeginExpansion
{\textstyle\bigcup\nolimits_{i=1}^{M}}
%EndExpansion
A_{i}$) there is no decoherence. Moreover, even if there is decoherence from
the viewpoint of all the $\mathcal{O}_{A_{i}}$, this does not imply
decoherence from the viewpoint of $\mathcal{O}_{A}$ since, as it is
well-known, $\mathcal{O}_{A}$ is not the mere union of the $\mathcal{O}%
_{A_{i}}\otimes\left(
%TCIMACRO{\tbigotimes \nolimits_{j=1,j\neq i}^{M}}%
%BeginExpansion
{\textstyle\bigotimes\nolimits_{j=1,j\neq i}^{M}}
%EndExpansion
I_{j}\right)  $. In other words, in agreement with quantum holism, the open
subsystem $A$ is not the mere collection of the particles $A_{i}$; then, it is
reasonable to expect that the behavior of $A$ cannot be inferred from the
behavior of all the $A_{i}$. In the same sense, it is not surprising that
there is no decoherence from the viewpoint of the total space of observables
$\mathcal{O}$ of $U$, in spite of the fact that there is decoherence from the
viewpoint of anyone of the $\mathcal{O}_{A_{i}}$ and $\mathcal{O}_{B_{k}}$,
corresponding to the particles $A_{i}$ and $B_{k}$ respectively. And since the
privileged viewpoint does not exist, the conclusions about decoherence have to
be relativized to the particular observational perspective selected in each case.

\section{A new look at environment-induced decoherence}

\subsection{Dissolving the conceptual problems}

When decoherence is understood in the new general framework, the conceptual
difficulties of the EID program turn out to be not as serious as originally
supposed. In fact:\bigskip

\begin{enumerate}
\item[a)] With respect to the \textquotedblleft open-system\textquotedblright%
\ problem, it does not make sense to say that closed quantum systems may not
decohere: a closed quantum system may decohere from certain observational
perspectives, that is, from the viewpoint of certain spaces of observables.
Furthermore, in spite of the fact that EID focuses on open systems, it can
also be formulated from the perspective of the composite closed system and, in
this case, meaningful relationships between the behavior of the whole system
and the behavior of its subsystems can be explained (see
\cite{Castagnino-Laura-Lombardi-2007}, \cite{General Framework}).

\item[b)] The \textquotedblleft defining systems\textquotedblright\ problem is
simply dissolved by the fact that the splitting of the closed system into an
open subsystem and an environment is just a way of selecting the relevant
observables of the closed system. Since there are many different sets of
relevant observables depending on the observational viewpoint adopted, the
same closed system can be decomposed in many different ways: each
decomposition represents a decision about which degrees of freedom are
relevant and which can be disregarded in any case. Since there is no
privileged or \textquotedblleft essential\textquotedblright\ decomposition,
there is no need of an unequivocal criterion for deciding where to place the
cut between \textquotedblleft the\textquotedblright\ system and
\textquotedblleft the\textquotedblright\ environment. As said above,
decoherence is not a yes-or-no process, but a phenomenon relative to the
chosen decomposition. Therefore, Zurek's \textquotedblleft looming big
problem\textquotedblright\ is not a real threat to the EID approach: the
supposed challenge dissolves once the relative nature of decoherence is taken
into account (see \cite{Castagnino-Fortin-Lombardi-2010b},
\cite{Lombardi-Fortin-Castagnino-2011}).
\end{enumerate}

Although the new framework neutralizes the conceptual difficulties of the EID
approach, also points to some warnings about the way in which the proposal is
usually presented. From the new perspective, the insistence on the essential
role played by the openness of a system and its interaction with the
environment in the phenomenon of decoherence sounds rather misleading. The
essential physical fact is that, among all the observational viewpoints that
may be adopted to study a closed system, some of them determine a subset of
relevant observables for which the system decoheres.

\subsection{Decoherence and dissipation}

As pointed out in Section II, certain presentations of the EID approach
suggest the existence of a certain relationship between decoherence and
dissipation, as if decoherence were a physical consequence of or, at least,
were strongly linked to energy dissipation. Some particular models studied in
the literature on the subject tend to reinforce this idea by describing the
behavior of a small open system $-$typically, a particle$-$ immersed in a
large environmental bath. On this basis, the EID approach has been considered
a \textquotedblleft dissipative\textquotedblright\ approach, by contrast to
\textquotedblleft non-dissipative\textquotedblright\ accounts of decoherence
that constitute the \textquotedblleft heterodoxy\textquotedblright\ in the
field (\cite{Bonifacio}, \cite{Ford}, \cite{Frasca}, \cite{Sicardi},
\cite{Gambini}).

The fact that energy dissipation is not a condition for decoherence has been
clearly emphasized by Schlosshauer, who claims that \textquotedblleft%
\textit{decoherence may, but does not have to, be accompanied by dissipation,
whereas the presence of dissipation also implies the occurrence of
decoherence}\textquotedblright\ (\cite{Schloss-2007}, p.93). This fact is
explained by stressing that the loss of energy from the system is a classical
effect, leading to thermal equilibrium in the relaxation time, whereas
decoherence is a pure quantum effect that takes place in the decoherence time,
many orders of magnitude shorter than the relaxation time: \textquotedblleft%
\textit{If dissipation and decoherence are both present, they are usually
quite easily distinguished because of their very different timescales}%
\textquotedblright\ (\cite{Schloss-2007}, p.93). According to the author, it
is this crucial difference between relaxation and decoherence timescales that
explains why we observe macroscopic objects to follow Newtonian trajectories
$-$effectively \textquotedblleft created\textquotedblright\ through the action
of decoherence$-$ with no manifestation of energy dissipation, such as a
slowing-down of the object. Schlosshauer recalls an example used by Joos
(\cite{Joos et al}): the planet Jupiter has been revolving around the sun on a
Newtonian trajectory for billions of years, while its motional state has
remained virtually unaffected by any dissipative loss.

This explanation, although correctly stressing the difference between
decoherence and dissipation, seems to present both phenomena on the same
footing: an open system would first become classical through decoherence, and
would then relax due to energy dissipation. Following this view, dissipation
involves the loss of energy from the system to the environment, while
decoherence amounts to a sort of \textquotedblleft
dissipation\textquotedblright\ of coherence that leads the open system, in a
very short time, to the classical regime: the environment plays the role of a
\textquotedblleft sink\textquotedblright\ that carries away the information
about the system (\cite{Schloss-2007}, p.85). The results obtained in the
generalized spin-bath model show that the coherence-dissipation or
information-dissipation picture has to be considered with great caution, as a
mere metaphor. In fact, to the extent that decoherence is a relative
phenomenon, no flow of a non-relative quantity from the open system to the
environment can account for decoherence. In particular, although energy
dissipation and decoherence are in general easily distinguished because of
their different timescales, the very reason for their difference is that
energy dissipation is not a relative phenomenon but results from the effective
flow of a physical entity, whereas decoherence is relative to the
observational partition of the whole closed system selected in each situation.
On the other hand, decoherence can be explained in terms of the flow of
information from the open system to the environment if information is also
conceived as a relative magnitude (\cite{Lombardi-2004}, \cite{Lombardi-2005}).

\section{The quantum-to-classical transition}

As it is well known, one of the main aims of the decoherence program is to
explain \textquotedblleft\textit{the transition from quantum to classical}%
\textquotedblright\ (\cite{Zurek-1991}, p.36). For this purpose, it is
expected that the phenomenon of decoherence cancels the quantum features that
preclude classicality. In this section we will see how this may happen.

Let us begin by recalling that, in the quantum domain, the expectation values
are the experimentally accessible magnitudes: we do not measure states but
expectation values. Therefore, the purpose must be to obtain quantum
expectation values with a classical-like form. The expectation value of a
classical observable $O$, whose possible values are $o_{i}$, is computed as%

\begin{equation}
\langle O\rangle=\sum_{i}P_{i}\,o_{i} \label{37}%
\end{equation}
where $P_{i}$ is the probability corresponding to the value $o_{i}$. In
quantum mechanics, by contrast, the expectation value of an observable $O$ in
the state $\rho$ is computed as%

\begin{equation}
\langle O\rangle_{\rho}=Tr\left(  \rho O\right)  =\sum_{ij}\rho_{ji}%
\,O_{ij}\,=\sum_{i}\rho_{ii}\,O_{ii}+\sum_{i\neq j}\rho_{ji}\,O_{ij}=%
%TCIMACRO{\tsum \nolimits^{D}}%
%BeginExpansion
{\textstyle\sum\nolimits^{D}}
%EndExpansion
+%
%TCIMACRO{\tsum \nolimits^{ND} }%
%BeginExpansion
{\textstyle\sum\nolimits^{ND}}
%EndExpansion
\label{38}%
\end{equation}
where the $O_{ij}$ and the $\rho_{ji}$ are the components of $O$ and $\rho$ in
a given basis. It is important to note that in this approach, the terms of the
mean values must be written in the same preferred basis of EID (however it is
also possible to define the baseline from the mean values \cite{MPB1}
\cite{MPB2}). Therefore, if the purpose is that quantum expectation values
behave as in the classical case, we need to identify the situation in which
they acquire a form that resembles the classical one, that is, a situation in
which the \textquotedblleft non-diagonal\textquotedblright\ sum $%
%TCIMACRO{\tsum \nolimits^{ND}}%
%BeginExpansion
{\textstyle\sum\nolimits^{ND}}
%EndExpansion
$ vanishes and only the \textquotedblleft diagonal\textquotedblright\ sum $%
%TCIMACRO{\tsum \nolimits^{D}}%
%BeginExpansion
{\textstyle\sum\nolimits^{D}}
%EndExpansion
$ remains.

\subsection{The quantum-to-classical transition in environment-induced
decoherence}

Since the non-diagonal terms of a quantum state do not have a classical
analogue, in the EID approach the diagonalization of the reduced state of the
system is viewed as a manifestation of the quantum-to-classical transition:
the states of the pointer basis, \textquotedblleft
einselected\textquotedblright\ by decoherence, are the candidates to classical
states (strictly speaking, in order to obtain a classical-like description it
is necessary to apply the Wigner transformation and the limit $\hbar
\longrightarrow0$, see \cite{Paz-Zurek-2002}). In most models, the system is a
particle and, as a consequence, it is usually said that the particle becomes
$-$or, at least, begins to behave as if it were$-$ classical.

In fact, the diagonalization of the reduced state $\rho_{S}(t)$ of $S$ in the
pointer basis leads to a situation in which certain expectation values acquire
a classical-like form. Since $\rho_{S}^{d}$ is diagonal, it can be conceived
as representing a classical-like probability that makes the interference terms
represented by $%
%TCIMACRO{\tsum \nolimits^{ND}}%
%BeginExpansion
{\textstyle\sum\nolimits^{ND}}
%EndExpansion
$ to be zero (see eq.(\ref{9})):%

\begin{equation}
\left\langle O_{S}\right\rangle _{\rho_{S}^{d}}=\sum_{i}\rho_{S{}ii}%
^{d}\,O_{S{}ii} \label{39}%
\end{equation}
As a consequence, the expectation value of any observable $O_{S}^{c}$ of $S$
that is diagonal in the pointer basis (the same of EID) will have the
classical form of eq.(\ref{37}):%

\begin{equation}
\left\langle O_{S}^{c}\right\rangle _{\rho_{S}^{d}}=\sum_{i}\rho_{S{}ii}%
^{d}\,o_{S{}i}^{c} \label{40}%
\end{equation}
where the diagonal components $\rho_{S{}ii}^{d}$ of $\rho_{S}^{d}$ play the
role of the classical-like probabilities corresponding to the possible values
$-$eigenvalues$-$ $o_{S{}i}^{c}$ of the observable $O_{S}^{c}$.

In summary, the decoherence of the observables corresponding to the system $S$
selects the pointer basis in which the reduced state $\rho_{S}(t)$ becomes
diagonal after the decoherence time and, in turn, that pointer basis
identifies the observables of $S$ whose expectation values acquire a classical
form. As a consequence, instead of saying that the system $S$ behaves as if it
were classical or that the states of the pointer basis become classical, it
would be more accurate to say that \textit{the observables} $O_{S}^{c}$ of $S$
behave as if they were classical observables. Moreover, on the basis of the
arguments of the previous sections, from the closed-system perspective one
could also say that the observables $O_{R}^{c}=O_{S}^{c}\otimes I_{E}$ of the
whole closed system $U=S\cup E$ behave as if they were classical observables.

\subsection{The quantum-to-classical transition: a more general approach}

As we have said in Section VI, from our viewpoint decoherence is a process
that leads $\left\langle O_{R}\right\rangle _{\rho_{U}(t)}$ to converge, in an
extremely short time, to a value that can be computed as if the system were in
a state represented by a diagonal density operator (see eq.(\ref{16})):%

\begin{equation}
\left\langle O_{R}\right\rangle _{\rho_{U}^{d}}=\sum_{i}\rho_{Uii}^{d}%
\,O_{R{}ii} \label{41}%
\end{equation}
Now $-$analogously to the EID case$-$ the expectation value of any observable
$O_{R}^{c}$ of $U$\ that is diagonal in the pointer basis will have the
classical form of eq.(\ref{37}):%

\begin{equation}
\left\langle O_{R}^{c}\right\rangle _{\rho_{U}^{d}}=\sum_{i}\rho_{Uii}%
^{d}\,o_{R{}i}^{c} \label{42}%
\end{equation}
where the diagonal components $\rho_{Uii}^{d}$ of $\rho_{U}^{d}$ play the role
of the classical-like probabilities corresponding to the possible values
$-$eigenvalues$-$ $o_{R{}i}^{c}$ of the observable $O_{R}^{c}$.

In complete analogy with the quantum-to-classical transition in EID, here the
decoherence of the relevant observables of $U$ selects the pointer basis in
which the state $\rho_{U}^{d}$ is diagonal and, in turn, this pointer basis
identifies, among all the relevant observables, those whose expectation values
acquire a classical form. But in this case, the relevant observables do not
need to correspond to a subsystem of $U$: it is possible to focus attention on
any group of observables of the closed system, or even on a single observable,
in order to know if they behave in a classical-like manner.

Since according to the EID approach only open systems may decohere, from this
perspective the issue of the emergence of classicality in closed systems, in
particular, in the Universe as a whole, cannot even be posed since
\textquotedblleft\textit{the state of a perfectly isolated fragment of the
Universe }$-$\textit{of, for that matter, of the quantum universe as a
whole}$-$\textit{\ would evolve forever deterministically}\textquotedblright%
\ (\cite{Zurek-1994}, p.181). Nevertheless, this difficulty can be overcome
from the closed-system perspective. In fact, once the idea of the decoherence
and the classicality of systems or states is replaced by the view that certain
observables decohere and may acquire a classical-like behavior, the
restriction to open systems loses its strength.

Moreover, the perplexities resulting from the generalized spin-bath model can
also be solved. In fact, they derive from supposing that the open subsystems
become classical when interacting with their environments. From the
open-system perspective, on the contrary, one can univocally identify, among
all the observables of the closed system, which acquire a classical-like
behavior: the \textquotedblleft classical world\textquotedblright\ is
unambiguously defined by the set observables whose expectation values acquire
a classical form.

\section{Decoherence without decoherence}

Current works indicate that the most suitable scenario for early cosmology is
the one of inflation \cite{kolb,Mukahnov,Peacock,Weinberg,coles,peebles}. This
is because the whole structure of the universe can be traced to the primordial
fluctuations during an accelerated phase of the early universe. Several works
show that EID can solve the quantum-classical transition of the universe, by
separating the different degrees of freedom in system and environment. Since
there is no a standard way to divide the universe, some authors take the
inflaton as a system and the gravitons as the environment
\cite{FrancoCalzetta}, while other authors divide the universe between long
and short frequencies \cite{Lombardo}.

However there is an alternative approach proposed by Kiefer, Polarski and
Starobinsky \cite{Starobinsky,Polarski,Kiefer}, it studies the problem of
quantum-classical transition without the need to attend to decoherence. The
predictions obtained through mean values are indistinguishable from stochastic averaging.

According to paper \cite{Kiefer}, inflation can be appears in energy scales
where spacetime can be described as a classic curved spacetime where
fluctuations are defined. Inflaton fluctuations $\delta\phi(x,t)$ can be
studied as scalar fields without mass. It is convenient to consider the
rescaled quantity $a\delta\phi(x,t)\equiv y(x,t)$ and the conformal time
$\eta=%
%TCIMACRO{\dint \limits_{{}}^{{}}}%
%BeginExpansion
{\displaystyle\int\limits_{{}}^{{}}}
%EndExpansion
\frac{dt}{a(t)}$. The symbol $^{\prime}$ will be used to denote a derivative
with respect to $\eta$. The quantization of the perturbation $y(x,\eta)$ is
the usual quantization, and $p$ is the momentum conjugate to $y$,%

\begin{equation}
p=\frac{\partial\mathcal{L}(y,y^{\prime})}{\partial y^{\prime}} \label{dsd2}%
\end{equation}
where $\mathcal{L}$\ is the Lagrangian that corresponds to the Hamiltonian:%

\begin{equation}
H=\frac{1}{2}%
%TCIMACRO{\dint \limits_{{}}^{{}}}%
%BeginExpansion
{\displaystyle\int\limits_{{}}^{{}}}
%EndExpansion
d^{3}k[k(a(k)a^{\dag}(k)+a^{\dag}(-k)a(-k))+i\frac{a^{\prime}}{a}(a^{\dag
}(k)a^{\dag}(-k)-a(k)a(-k)] \label{dsd5}%
\end{equation}

$a(k)$ is the time dependent annihilation operator%
\begin{equation}
a(k)=\frac{1}{\sqrt{2}}(\sqrt{k}y(k)+\frac{i}{\sqrt{k}}p(k)) \label{dsd6}%
\end{equation}

$y(k,\eta)$ and $p(k^{\prime},\eta)$ satisfy the canonical commutation relation%

\begin{equation}
\left[  y(k,\eta),p(k^{\prime},\eta)\right]  =i\delta^{(3)}(k-k^{\prime})
\label{dsd4}%
\end{equation}
This equation expresses the quantum character of the perturbation $y(k,\eta)$.

If we compute the time evolution in the Heisenberg representation we have that
the following:%

\begin{equation}
y(k,\eta)=f_{k}(\eta)a_{k}+f_{k}^{\ast}(\eta)a_{-k}^{\dag} \label{dsd11}%
\end{equation}
where $a_{k}=a(k,\eta_{0})$ and the modes $f_{k}$ satisfy the equation%

\begin{equation}
f_{k}^{\prime\prime}+(k^{2}-\frac{a^{\prime\prime}}{a})f_{k}=0 \label{dsd3}%
\end{equation}
We can write the equation (\ref{dsd11}) in a more suggestive way%

\begin{equation}
y(k,\eta)=\sqrt{2k}f_{k1}(\eta)y_{k}-\sqrt{\frac{2}{k}}f_{k2}(\eta)p_{k}
\label{dsd12}%
\end{equation}
where $y_{k}=y(k,\eta_{0})$ and $p_{k}=p(k,\eta_{0})$, $f_{k1}%
=\operatorname{Re}(f_{k})$, $f_{k2}=\operatorname{Im}(f_{k})$. Analogously%

\begin{equation}
p(k,\eta)=\sqrt{\frac{2}{k}}g_{k1}(\eta)p_{k}+\sqrt{2k}g_{k2}(\eta)y_{k}
\label{dsd13}%
\end{equation}

\subsection{Quantum-classical transition: pragmatic version}

This formulation is relevant to understand the quantum-classical transition.
For example, Kiefer and Polarski \cite{K-P} assume that there is a limit in
which $f_{k2}$ and $g_{k1}$ (or $f_{k1}$ and $g_{k2}$ respectively) behave as%
\begin{align*}
f_{k2}  &  \longrightarrow0\text{ \ \ \ \ and \ \ \ \ }g_{k1}\longrightarrow
0\\
&  \text{or}\\
f_{k1}  &  \longrightarrow0\text{ \ \ \ \ and \ \ \ \ }g_{k2}\longrightarrow0
\end{align*}

Then from (\ref{dsd12}) and (\ref{dsd13}) we can see that the
non-commutativity of operators $y_{k}$ and $p_{k}$ is no longer relevant%
\begin{equation}
\left[  y(k,\eta),p^{\dag}(k^{\prime},\eta)\right]  =2\left(  f_{k1}%
(\eta)g_{k1}(\eta)+f_{k2}(\eta)g_{k2}(\eta)\right)  \left[  y_{k}%
,p_{k}\right]  \longrightarrow0 \label{dsd13b}%
\end{equation}
To understand the physical meaning of this limit, we can consider a classical
stochastic system where the dynamics is still described by equations of the
form (\ref{dsd12}), but with $y(k,\eta_{0})$ and $p(k,\eta_{0})$ representing
random initial values. If $f_{k2}$ and $g_{k1}$ vanish, we get%

\begin{equation}
p(k,\eta)=p_{cl}(y(k,\eta))=\frac{g_{k2}}{f_{k1}}y(k,\eta) \label{dsd14}%
\end{equation}

This equation is valid when $y$ and $p$ are operators and also if they are
classical stochastic variables ($y_{cl}$ and $p_{cl}$). In this case, for a
given realization there are a pair $\left(  y_{cl},p_{cl}\right)  $. For this
reason Kiefer and Polarski claim

\textquotedblleft\textit{Then the quantum system is effectively equivalent to
the classical random system, which is an ensemble of classical trajectories
with a certain probability associated to each of them\textquotedblright}
(\cite{Kiefer} pp. 4)

A concrete example in which $f_{k2}$ and $g_{k1}$ tend to zero, is the
perturbation on de Sitter space \cite{K-P}. When the decaying mode becomes
vanishingly small; when this mode is neglected we are in the limit of a random
stochastic process.

\subsection{Pragmatic view, EID and the closed-system perspective}

Kiefer and Polarski present their explanation of the quantum-classical
transition of the universe as a \textquotedblleft pragmatic
view\textquotedblright. This is because although we can obtain classical
statistics from quantum statistics, there is not an associated interpretation.
The main characteristic of the \textquotedblleft pragmatic
view\textquotedblright\ is that it focuses on the mean values and it does not
use any reduced state.

Most interpretations of quantum mechanics emphasize the quantum state,
assigning to it some kind of ontology. In fact, in EID the focus is in the
reduced state. For this reason the authors believe that their point of view,
that does not consider the states, is simply a pragmatic view. And to
understand the phenomenon is necessary to introduce EID.

However, we believe that this formalism can be given in a proper interpretive
framework that does not consider to the state, then EID is not necessary

The pragmatic view basically consists of studying the mean values {}%
{}directly, without taking into account the state, and it determines when
these values {}{}do take the form of classical statistics and when they do
not. This way of studying decoherence is similar to the approach proposed by
the closed-system perspective.

\subsubsection{The hidden decoherence of the \textquotedblleft pragmatic
view\textquotedblright}

As it is known, a quantum system evolves under the Schr\"{o}dinger equation.
As the evolution is unitary, the system cannot reach equilibrium or decoherence.

As explained in previous sections, if we choose appropriate relevant
observables we can describe the phenomenon of decoherence in the mean values
of the closed system. Therefore, the quantum-classical transition of the
\textquotedblleft pragmatic view\textquotedblright\ has to ignore some kind of
decoherence. This is difficult to determine if the decoherence is understood
from the point of view of EID, since it divided the whole system into the
system and the environment, computes the reduced state of the system and
considers its diagonalization. However, according to the closed-system
perspective, decoherence is understood as the vanish of the interference terms
in the mean values, then the problem is simplified. In fact,the pragmatic view
selects the observable ad ignores evanescent modes.

The expression (\ref{dsd13b}) is misleading because from (\ref{dsd4}),
(\ref{dsd12}) and (\ref{dsd13}) we know that%
\begin{equation}
f_{k1}(\eta)g_{k1}(\eta)+f_{k2}(\eta)g_{k2}(\eta)=\frac{1}{2}%
\end{equation}
then
\begin{equation}
\left[  y(k,\eta),p^{\dag}(k^{\prime},\eta)\right]  \nrightarrow0
\end{equation}

Then Kiefer and Polarski say

\begin{center}
\textquotedblleft\textit{Clearly, when }$f_{k2}(\eta)$\textit{, }$g_{k1}%
(\eta)$\textit{ are unobservable, this coherence becomes unobservable as well.
This is the case when the decaying mode is so small that we have no access to
it in observations\textquotedblright\ }(\cite{Kiefer} pp. 5)
\end{center}

This means that coherence becomes unobservable from the observational point of
view, as in the closed-system perspective. Since $f_{k2}$ and $g_{k1}$ are
undetectable, it requires a gross observable that is not capable to detect the
presence of $f_{k2}$ and $g_{k1}$. That is, we need to select a relevant
observable $O_{R}$. If, for example, we have only access to the observable%
\begin{equation}
O_{R}=\left(  \left\vert f_{k2}(\eta)\right\vert +\left\vert g_{k1}%
(\eta)\right\vert \right)  I
\end{equation}
then%
\begin{equation}
\left\langle O_{R}\right\rangle \longrightarrow0
\end{equation}
In this case we can say that the system decoheres from the observational point
of view. Then the pragmatic view of Kiefer, Polarski and Starobinsky is fully
compatible with the closed-system perspective.

\section{\textbf{Conclusions}}

The aim of this paper has been to argue that decoherence can be viewed from a
closed-system perspective, which improves the understanding of the phenomenon.
After recalling the historical development of the decoherence program and
stressing some conceptual difficulties of the EID program, we have introduced
the closed-system approach from which the orthodox EID approach can be
reformulated. In this theoretical framework we have analyzed the spin-bath
model, both in its traditional and in a generalized version: by considering
different partitions of the whole closed system in both cases, we have shown
that decoherence depends on the way in which the relevant observables are
selected. On this basis, the following conclusions can be drawn:

\begin{enumerate}
\item[(i)] Decoherence is a phenomenon relative to which degrees of freedom of
the whole closed system are considered relevant and which are disregarded in
each situation.

\item[(ii)] Since there is no privileged or essential decomposition of the
closed system, there is no need of an unequivocal criterion for identifying
the systems involved in decoherence. Therefore, the \textquotedblleft looming
big problem\textquotedblright\ $-$which, according to Zurek, poses a serious
threat to the whole decoherence program$-$ dissolves in the light of the
relativity of decoherence.

\item[(iii)] Due to its relative nature, decoherence cannot be accounted for
in terms of dissipation of energy or of any other non-relative magnitude.

\item[(iv)] The open-system perspective leads to the unambiguous
identification, among all the observables of the closed system, of those that
will have a classical-like behavior because their expectation values acquire a
classical form.
\end{enumerate}

Once the phenomenon of decoherence is \textquotedblleft
de-substantialized\textquotedblright\ in this way, one might ask in what sense
it can be still understood as the result of the action of an environment that
destroys the coherence between the states of an open quantum system.

\section{\textbf{Acknowledgments}}

We are very grateful to Roland Omn\`{e}s and Maximilian Schlosshauer for many
comments and criticisms. This research was partially supported by grants of
the University of Buenos Aires, CONICET and FONCYT of Argentina.

\end{document}